\documentclass[11pt]{article}

\usepackage[preprint]{acl}

\usepackage{times}
\usepackage{latexsym}
\usepackage{caption}
\usepackage{hyperref}

\usepackage[T1]{fontenc}
\usepackage[utf8]{inputenc}

\usepackage{microtype}

\usepackage{inconsolata}

\usepackage{graphicx}
\usepackage{booktabs}
\usepackage{array}
\usepackage{multirow}
\usepackage{listings}
\usepackage{subcaption}  % 支持子标题
\usepackage{courier}
\usepackage{xcolor}
\usepackage[most]{tcolorbox}
\usepackage{adjustbox}
\title{VIVID: Bringing Tool-Augmented Agents into Enterprises}

\author{
 \textbf{Yichu Fang},
 \textbf{Ziqiang Zhang},
 \textbf{Qiyang Shi},
 \textbf{Zilong Wang},
 \textbf{Lin Shen}, \\
 \textbf{Zhikang Ou},
 \textbf{Jing Ma},
 \textbf{Jiayuan Chen},
 \textbf{Lei Sun},
 \textbf{Dai Cheng},\\
 \textbf{Yi Qiao},
 \textbf{Yu He},
 \textbf{Wei Zhang},
 \textbf{Xiaoyu Shen\thanks{Corresponding author}}
\\
 Ningbo Institute of Digital Twin, Eastern Institute of Technology, Ningbo\\
 \href{mailto:xyshen@eitech.edu.cn}{xyshen@eitech.edu.cn}
}

\begin{document}
\maketitle
\begin{abstract}
Tool-augmented large language model agents connect natural-language requests to enterprise services, yet their operational viability depends on reliable execution, organizational control, and system responsiveness. We present VIVID, a deployed tourism-management agent system designed to study these requirements in a real enterprise environment. On an audited dataset of 283 queries, its Qwen3-8B configuration achieves 59.06\% strict end-to-end success across the evaluated decision stages, indicating useful but bounded autonomous capability. Stage-wise KV-cache eviction reduces mean latency by 34.4\%, while targeted optimization reduces the mean latency of the five slowest measured business tools by 81.1--92.3\%.  In a three-month deployment serving more than 1,500 personnel, VIVID reduced average per-order processing time from approximately ten minutes to under two minutes, with a final-task error rate below 1\% after human-in-the-loop control. These results show that enterprise agent viability requires system-level consideration beyond isolated tool-calling accuracy.
\end{abstract}

\section{Introduction}
Tool-augmented large language model (LLM) agents offer a promising interface for enterprise software by translating natural-language goals into tool invocations. Recent advances in agentic reasoning, API-oriented instruction tuning, large-scale tool retrieval, and standardized interfaces have substantially improved this capability~\cite{yao2022react,schick2023toolformer,liu2025toolace,shi2025retrieval}. Benchmark performance, however, does not establish whether a complete agent system can operate under enterprise constraints. Consider a tourism operator using an embedded agent on an order-management page to complete an administrative approval. The agent can reduce the manual effort required to locate and coordinate relevant tools, but it must work with proprietary, heterogeneous, and continually evolving tools. Its actions must follow organizational authorization, remain visible to users, and complete with acceptable latency across model inference and external execution. Incorrect actions may affect business operations, whereas slow responses can limit adoption. We therefore investigate the operational viability of a complete tool-augmented agent system under enterprise reliability, governance, and latency constraints.

Operational viability depends on three connected challenges. First, errors can accumulate as a request passes through multiple agentic interaction steps, limiting the reliability of fully autonomous execution. At the same time, probabilistic model decisions cannot substitute for organizational policy: consequential actions require authorization enforced outside the model and mechanisms for retaining human authority for controllability~\cite{amershi2019guidelines,buccinca2021trust}. Responsiveness presents a further constraint, as iterative inference introduces model-side latency while concurrent business operations impose additional service-side overhead. These challenges together determine whether agentic automation can reduce the cognitive load of coordinating multi-step business tasks for employees without compromising operational control and responsiveness~\cite{sweller1988cognitive,prabhudesai2026evaluating}.

A city-scale tourism management system provides a realistic setting for this investigation. Its business tasks span multiple departments, data sources, and services. Existing platforms typically expose operations through predefined interfaces and fixed business logic, providing stability for standardized tasks while requiring staff to translate high-level business intents into subsystem-specific actions~\cite{lam2009travel}. Cross-functional tasks and business rules increase this coordination burden and limit the flexibility of conventional interfaces~\cite{liu2025pricinglogic}. The setting therefore exposes both the potential value of a natural-language entry point and the operational constraints that surround executable actions.

We present \textsc{VIVID}, a complete tool-augmented agent system deployed in a real-world production environment supporting city-scale enterprise tourism operations. The interactive interface provides the entry point for users to submit tasks and observe execution progress, while enabling intervention when needed. The Decision Core translates these requests into model-proposed actions grounded in the evolving enterprise tool inventory maintained by the Tool Management Hub, which supports tool onboarding and lifecycle management with MCP~\cite{anthropic2024mcp} as a standardized invocation interface. Before execution, the Application Control Gateway enforces authorization independently of model generation, ensuring that tool actions remain subject to organizational policy. To improve responsiveness, VIVID removes redundant inference state through stage-wise KV-cache eviction and combines runtime monitoring with targeted optimization to identify and mitigate bottlenecks in business services. Overall, these components support governed, observable, and latency-aware execution across the complete system.

Our contributions are threefold. (i) We present a governed agent-system architecture that integrates model-based action proposal, protected authorization, and visible, controllable execution. The architecture also provides lifecycle management for 73 deployed tools. (ii) We develop a full-path engineering approach that addresses latency in both iterative model inference and enterprise services through stage-wise cache eviction, runtime monitoring, and targeted optimization. (iii) We report complementary controlled and deployment evidence. On 283 audited queries, VIVID achieves 59.06\% strict end-to-end success across the evaluated decision stages. KV-cache eviction reduces mean latency by 34.4\% for tool-use tasks, and targeted optimization reduces the mean latency of the five slowest measured business tools by 81.1--92.3\%. During a three-month deployment serving more than 1,500 personnel, the system processed 2,375 orders with an observed final-task error rate below 1\% after the complete human-in-the-loop process; average per-order processing time decreased from approximately ten minutes to under two minutes, and 96.92\% of 1,300 voluntary and informed survey respondents reported improved work efficiency. The controlled results characterize technical capability and efficiency, while the observational deployment evidence establishes operational use at scale.

\section{Related Work}
\paragraph{Tool-Augmented LLM Systems}
Tool use extends LLMs beyond their parametric knowledge by enabling interaction with external environments. Prior work has studied API invocation, reasoning with environment feedback, and multi-tool execution~\cite{yao2022react,inaba2023multitool,schick2023toolformer}, followed by systems that cover larger API collections and more complex trajectories~\cite{qin2024toolllm,liu2025toolace,shi2025retrieval}. MCP provides a standardized interface for connecting AI systems with external tools~\cite{anthropic2024mcp}. However, these studies primarily evaluate model capabilities on benchmarks or in controlled environments, leaving how agents perform under the reliability, security, and operational constraints of real enterprise tasks largely unexplored.

\paragraph{KV Cache Management}
KV cache reduces repeated computation during LLM inference, while its memory demand grows with sequence length~\cite{li2024survey,shi2024keep}. Existing methods apply selective eviction, quantization, or low-rank compression at the token level~\cite{chen2024nacl,hooper2024kvquant,li2024survey}. Architectural and system techniques include grouped-query attention, hierarchical storage, and paged attention~\cite{ainslie2023gqa,liu2025lmcache,kwon2023efficient}. Adaptive compression and multi-tier storage further reduce memory use and inference latency~\cite{ge2023model,wang2026multi}. For tool-augmented systems, cache management also supports on-demand tool loading~\cite{jones2025code} and removal of redundant tool responses.

% Extended related-work notes retained from the original draft.
% MCP provides a standardized framework for connecting LLMs to external data sources and tools~\cite{hou2025model}. It addresses integration across AI applications and external systems~\cite{hasan2025model}. Prior studies have examined its architectural design and security implications~\cite{hou2025model,errico2025securing}. MCP-based code execution can also support on-demand tool loading, while MCP-Bench evaluates agents in multi-tool settings~\cite{jones2025code,wang2025mcp}. Enterprise-oriented discussions further describe protocol-level integration across models and platforms~\cite{narajala2025enterprise}.

\section{System Design}
\label{sec:03system}
% \begin{figure*}[htbp]
% \centering
%     \begin{subfigure}[t]{\textwidth}
%     \begin{minipage}[t]{\textwidth}
%     \centering
%         \includegraphics[width=\textwidth]{imgs/arch/arch_full.pdf}
%     \end{minipage}
%     \caption{System Overview}
%     \label{fig:03arch-full}
%     \end{subfigure}
% \hfill

%     \begin{subfigure}[t]{\textwidth}
%     \begin{minipage}[t]{\textwidth}
%     \centering
%         \includegraphics[width=\textwidth]{imgs/arch/arch_mcp.pdf}
%     \end{minipage}
%     \caption{}
%     \label{fig:03arch-mcp}
%     \end{subfigure}

% \caption{All-in-one system architecture}
% \label{fig:03-arch}
% \end{figure*}

\begin{figure*}[htbp]
    \centering
    \includegraphics[width=\textwidth]{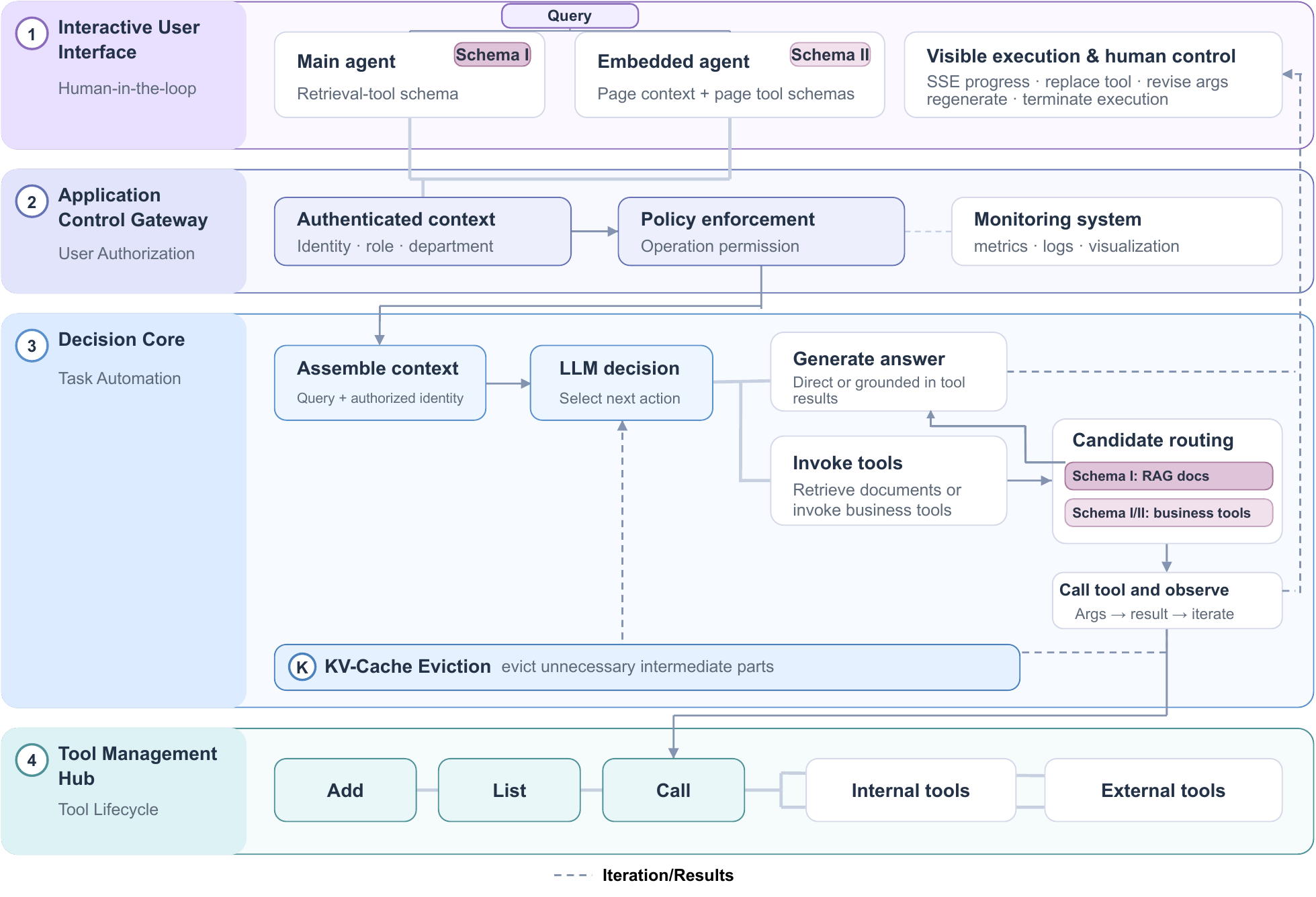}
    \caption{\small Agent system overview. Schema~I denotes the retrieval-tool schema; Schema~II represents the schema of page-specific business tools. Results from the backbone LLM are streamed to the interactive user interface. During the iterative agentic loop, intermediate tool-call records are evicted from the KV cache once their results have been incorporated into the workflow state.}
    \label{fig:03arch}
\end{figure*}

\subsection{Overview}
The goal of VIVID is to provide a natural-language entry point to enterprise applications, reducing the cognitive load associated with navigating heterogeneous interfaces and composing multi-step operations~\cite{sweller1988cognitive,prabhudesai2026evaluating}. VIVID can be described from two complementary perspectives. From the user-facing perspective, it operates as a tool-augmented agent that accepts a natural language request and returns a result, consistent with prior characterizations of LLM-based agents as systems that combine model decisions, actions, and environmental feedback~\cite{wang2024agentsurvey,yao2022react}. From the execution perspective, each interaction unfolds as an end-to-end workflow spanning the complete VIVID system. This workflow reconciles probabilistic model decisions with the need for governed and traceable enterprise execution while preserving human authority over consequential operations. Thus, \emph{agent} describes VIVID's externally observed behavior, whereas \emph{execution workflow} describes the governed system-wide process that realizes that behavior.

As shown in Figure~\ref{fig:03arch}, the execution workflow spans four components: the Interactive User Interface $\mathcal{I}$, Application Control Gateway $\mathcal{G}$, Decision Core $\mathcal{D}$, and Tool Management Hub $\mathcal{H}$. VIVID supports two agent modes. The main-agent mode handles general requests, and the embedded-agent mode retains these general capabilities while augmenting the initial model-visible state with the current page's task context and associated tool schemas, enabling assistance tailored to the ongoing task. For entry mode $e$, authenticated user $u$, and query $q$, the complete interaction is:
\begin{equation*}
\mathcal{W}_e(q,u)=\langle s_0,(a_t,o_t)_{t=0}^{T},r\rangle,
\label{eq:execution-workflow}
\end{equation*}
where $s_0$ is initialized from the query, protected user state, entry-specific context $c_e$, and available tool schemas $\mathcal{T}_e$. Each step derives an action $a_t$ from the current state, produces an observation $o_t$, and updates the state for the next step, with the final result $r$ returned to the user. Thus, $\mathcal{W}_e(q,u)$ represents the sequential execution trace of the agent.

Meanwhile, the Decision Core uses the LLM to propose $a_t$ based on the following elements:
\begin{equation*}
a_t=\pi_\theta(q,c_e,\mathcal{T}t,o_{<t}).
\end{equation*}
where $o_{<t}$ denotes observations from preceding steps. The proposed action then undergoes governed execution:

\begin{align*}
o_t &= \eta(s_t,a_t),\\
s_{t+1} &= \delta(s_t,a_t,o_t).
\end{align*}
where $\eta$ performs governed execution of $a_t$ under the protected state $s_t$, including authorization checks and potential human intervention, and produces $o_t$. Authorized tool executions are dispatched through standardized MCP interfaces~\cite{anthropic2024mcp}, while $\delta$ incorporates the observation into the protected state $s_{t+1}$. This separation distinguishes model-based action proposal from governed execution, preserving system-level control over model-generated actions.

VIVID also streams intermediate execution states and final responses through Server-Sent Events (SSE), providing users with visibility into ongoing execution and opportunities to intervene when necessary. This interaction model supports human-in-the-loop control, particularly for long-running, uncertain, or consequential operations~\cite{amershi2019guidelines,buccinca2021trust}.

\subsection{Bounded-Autonomy Interaction}
\label{sec:03-bounded}
Since model decisions can be incorrect and errors can accumulate across steps, VIVID bounds autonomy through organizational and interaction controls. Assuming a request originates from the embedded-agent mode on an order-management page, the Application Control Gateway validates the user role and injects the authorization context before execution, independently of the model generation. This ensures that access to the requested order is determined by the user's actual permissions rather than by model-generated arguments.

To support interaction controls, the interface exposes tool arguments, results, and elapsed time as the request advances through stages. A user can replace the selected tool with another retrieved candidate or revise task-specific parameters before continuing. The interface also supports response regeneration and termination of an active execution. System-managed authorization information remains inaccessible to both the user and the LLM, preserving the separation between model decisions and organizational control. Appendix~\ref{app:front} presents the complete SSE event lifecycle and detailed interface prototype design.

\subsection{Decision Core}
\label{sec:03-agent}
\paragraph{Agentic Loop}
The Decision Core provides the model-driven action policy for VIVID. It translates natural-language requests into executable actions by planning over the context and tools available at each $c_e$. All requests include the user query and authenticated identity information. The main-agent mode additionally receives the retrieval-tool schema, whereas an embedded-agent mode receives metadata about the current application page and its associated business tools. This module therefore supports both the general conversation and tool invocations.

As shown in Figure~\ref{fig:03arch}, for requests from the main agent (Schema~I), the Decision Core first invokes a retrieval tool to match the query against domain knowledge or candidate business tools. Only the retrieved candidates enter the prompt, limiting context expansion from the complete tool inventory. In contrast, requests from task-specific pages (Schema~II) directly incorporate the tools associated with that page, while the page description supports additional discovery when needed.

Following prior agentic work~\cite{yao2022react,qin2024toolllm}, the module executes a request iteratively. In Schema~I, the first step retrieves a ranked set of candidates, consisting either of relevant documents for retrieval-augmented generation (RAG) or candidate business tools. The LLM then either generates a response grounded in the retrieved documents or selects a business tool based on the user intent. For tool-based tasks, an additional selection step is performed over the candidate set to reduce retrieval uncertainty, and the selected tool is invoked in the subsequent step with model-generated arguments. Schema~II uses the same invocation mechanism with page-specific tools available from the beginning. Simple conversational requests can proceed directly to response generation.

\paragraph{Stage-Wise KV Cache Eviction}
Multi-tool agent execution can introduce substantial KV-cache redundancy, reducing system responsiveness. Existing approaches identify and retain salient components~\cite{tang2024quest,dehghanighobadi2026depthkv}, whereas multi-step agent interactions introduce stage-level redundancy when intermediate artifacts from a completed step become unnecessary after their information enters the workflow state $s_t$. VIVID therefore removes obsolete intermediate cache while retaining the context required for subsequent execution.

\begin{figure}[htbp]
    \centering
    \includegraphics[width=0.45\textwidth]{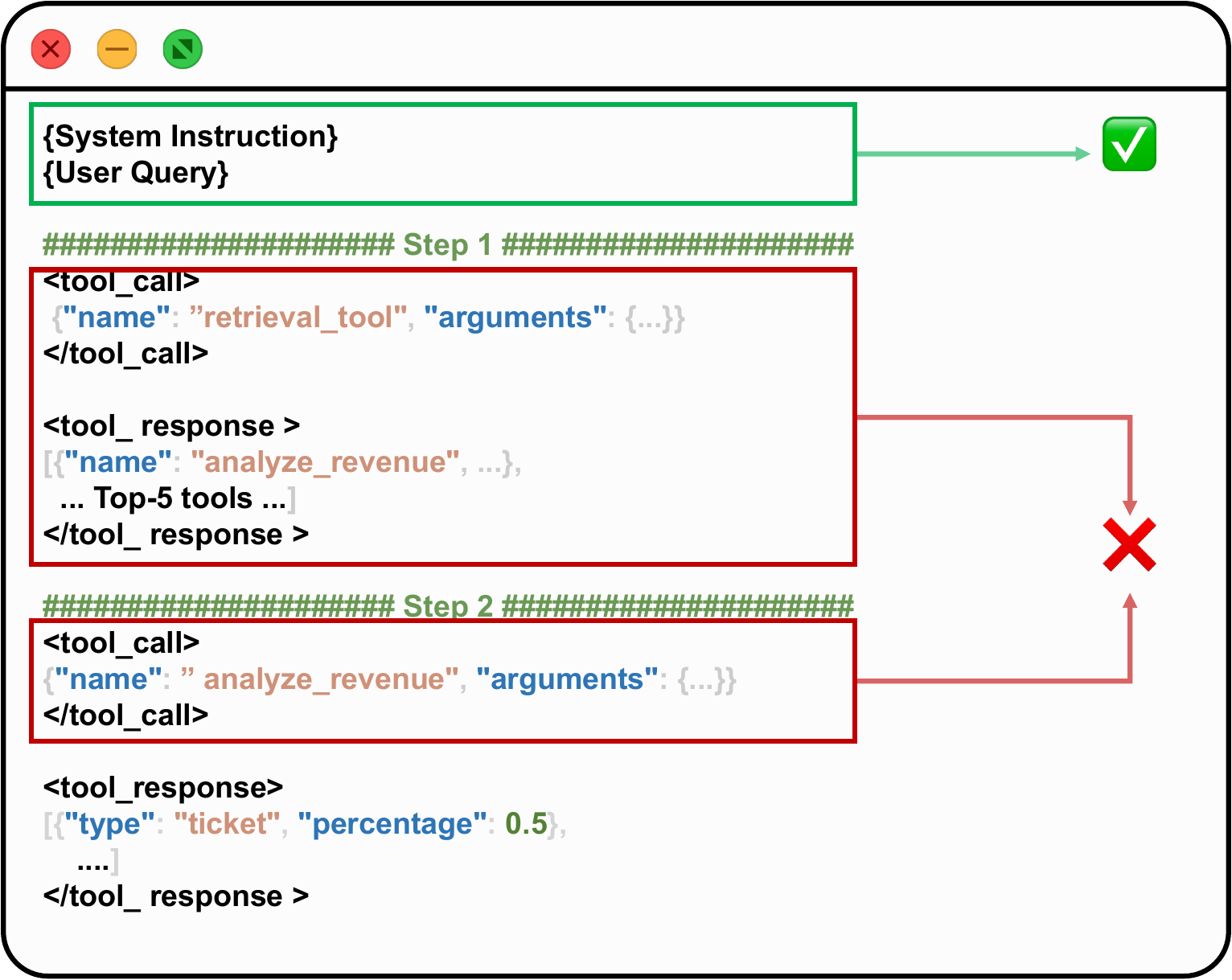}
    \caption{\small Example KV Cache Eviction for Step~3. The intermediate tool calls and responses are evicted.}
    \label{fig:03kv}
\end{figure}

Figure~\ref{fig:03kv} shows the prompt for summary generation at Step~3 in Schema~I. After the business tool is selected, the retrieval response and obsolete tool-call records are removed before the next step. The system instruction, user query, and informative tool response remain available for generation.
%The latency effect of this stage-wise policy is evaluated in Section~\ref{sec:efficiency}.

\subsection{Tool Management Hub}
\label{sec:03-hub}
\begin{figure}[htbp]
    \centering
    \includegraphics[width=0.48\textwidth]{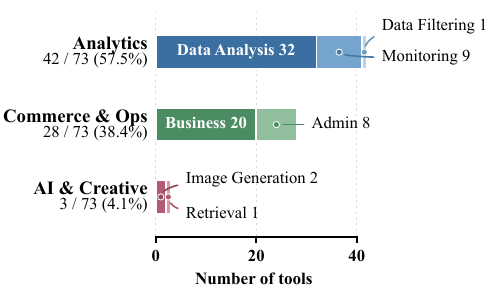}
    \caption{\small Distribution of tools across three primary categories and seven functional subcategories.}
    \label{fig:03tools}
\end{figure}

\paragraph{Tool Development}
Production deployment requires reliable, domain-specific tools. Open-source APIs and generated datasets provide broad coverage~\cite{liu2025toolace,shi2025retrieval,xu2025toucan}, while enterprise scenarios require schemas aligned with existing business services. Initial exploration found manual development inconsistent and labor-intensive to verify, while approximately 75\% of automatically generated candidate tools aligned with real business scenarios. VIVID therefore uses LLMs to draft schemas under a shared specification~\cite{guo2026learning,liu2025toolace,xu2025toucan} but retains developer verification over business semantics, implementation, and data exposure. Appendix~\ref{app:tool} provides details of the initial manual and automated development explorations. Developer-oriented identifiers are mapped to user-facing names without changing the identifiers used for model selection. Figure~\ref{fig:03tools} shows the 73 deployed tools. Analytical tools form the largest category, and business tools account for 71.4\% of Commerce \& Ops.

%Bar lengths represent category sizes, while colored segments and callouts report the corresponding subcategory counts.

\paragraph{Tool Management}
The Tool Management Hub maintains tools as business requirements evolve. \emph{Add} onboards schemas through the development pipeline, and \emph{List} provides access to registered definitions for inspection and reuse. \emph{Call} routes invocations to internal services or external providers according to each registered definition. MCP standardizes tool invocation, while the Tool Hub unifies the management of heterogeneous business tools and supports their lifecycle management. Tool definitions are distributed across storage partitions as the inventory grows, and concurrent invocations use managed connection pools.

\subsection{Application Control Gateway}
\label{sec:03-system}
\paragraph{Authorization Control}
Production tool use requires fine-grained access control across heterogeneous operations and user roles. VIVID centralizes authorization in the Application Control Gateway based on Casbin and Casdoor. Before a tool is invoked, the gateway enforces persisted policies by evaluating the authenticated user's role and department against the requested resource and operation. Centralized audit records provide a consistent basis for subsequent security monitoring.

\paragraph{Monitoring System}
Beyond stage-wise KV-cache eviction, iterative execution and concurrent requests require performance monitoring to identify bottlenecks and sustain production responsiveness. We deploy a monitoring stack based on Prometheus, Loki, and Grafana: Prometheus collects time-series metrics, Loki aggregates execution logs, and Grafana visualizes runtime behavior. The stack tracks latency, resource utilization, and invocation statistics for tool execution (\emph{Call}), tool listing (\emph{List}), and the chat entry-point API (\emph{Chat}). Stress tests at different concurrency levels guide optimization of SQL execution, connection management, and tool-set caching. Appendix~\ref{app:monitoring-diagnosis} presents the corresponding development-time diagnosis.

\section{Evaluation}
\subsection{Evaluation Setup}
\paragraph{Model Configuration}
The agent system uses Qwen3-8B to satisfy on-premises deployment requirements. Qwen3 provides a strong foundation for reasoning and agentic tasks~\cite{yang2025qwen3}, while VIVID requires supervision aligned with its business APIs and staged execution protocol. Prior studies similarly show the value of API-oriented training data for function calling~\cite{liu2025toolace} and task-specific adaptation for tool retrieval~\cite{shi2025retrieval}. We use the backbone as the feasibility baseline and train the production configuration on 1,601 task-specific examples for retrieval, tool selection, and argument generation.

\paragraph{Metrics}
We evaluate model effectiveness using four metrics. \textbf{Hit@5} measures whether the reference tool is retrieved among the top five candidates in Step~1 of Schema~I. \textbf{Selection accuracy} measures whether the Decision Core selects the reference business tool in Step~1 of Schema~II or Step~2 of Schema~I, conditioned on successful retrieval in Schema~I. \textbf{Argument accuracy} measures exact agreement between the generated and reference arguments after correct tool selection. \textbf{End-to-end success rate} (E2E SR) requires successful completion of every step for a query. We also assess system efficiency using execution latency metrics.

\subsection{Results}
\paragraph{Model Effectiveness}
The final audited dataset comprises 283 queries that are disjoint from the 1,601 training examples and cover the deployed tool inventory. The fine-tuned model achieves 88.77\% Hit@5, with conditional selection and argument accuracies of 82.45\% and 80.69\%, respectively, yielding an E2E SR of 59.06\%. This strict rate indicates useful but bounded one-execution capability over the audited scope. Errors accumulated across intermediate steps still constrain fully autonomous operation. The bounded-autonomy interaction therefore preserves visibility and user control for failed or operationally critical actions by design. Preliminary experiments also examine ambiguity in natural-language requests~\cite{budzianowski2018multiwoz,min2020ambigqa}. Explicit query rewriting is unnecessary in this setting, while overlapping tool capabilities require careful dataset verification. Appendix~\ref{app:initial} presents the retriever adaptation and auditing procedure.

\paragraph{System Efficiency}
\label{sec:efficiency}
We evaluate latency from both model and service perspectives. Across 20 tool-invocation requests, KV-cache eviction reduces mean latency from 5,827.57 to 3,820.45~ms (34.4\%), and P95 latency from 10,242.77 to 7,517.72~ms (26.6\%). Additionally, Table~\ref{tab:efficiency} compares the average execution latency of the five slowest business tools before and after the combined system optimizations, capturing the high-latency tail of the inventory. Their mean execution latency decreases by 81.1--92.3\%, with the slowest tool improving by more than three seconds.

\begin{table}[ht]
\centering
\small
\setlength{\tabcolsep}{4pt}
\renewcommand{\arraystretch}{0.9}
\begin{tabular}{@{}lcc@{}}
\toprule
\textbf{Tool} & \textbf{Latency (ms)} & \textbf{Reduction (\%)} \\
\midrule
\textit{analyze\_resource\_usage}
    & 3656 $\rightarrow$ \textbf{526} & 85.6 \\
\textit{analyze\_resort\_revenue}
    & 3397 $\rightarrow$ \textbf{429} & 87.4 \\
\textit{analyze\_team\_scales}
    & 1973 $\rightarrow$ \textbf{372} & 81.1 \\
\textit{analyze\_associations}
    & 1198 $\rightarrow$ \textbf{92} & 92.3 \\
\textit{analyze\_regional\_data}
    & 827 $\rightarrow$ \textbf{97} & 88.3 \\
\bottomrule
\end{tabular}
\caption{\small Tool latency before and after optimization.}
\label{tab:efficiency}
\end{table}

% \begin{table}[ht]
% \centering
% \small
% \begin{tabular}{lccc}
% \toprule
% \textbf{Tool} & \textbf{Before (ms)} & \textbf{After (ms)} \\
% \midrule
% \textit{analyze\_resource\_usage} & 3656 & \textbf{526} \\
% \textit{analyze\_resort\_revenue} & 3397 & \textbf{429} \\
% \textit{analyze\_team\_scales} & 1973 & \textbf{372} \\
% \textit{analyze\_associations} & 1198 & \textbf{92} \\
% \textit{analyze\_regional\_data }& 827 & \textbf{97} \\
% \bottomrule
% \end{tabular}
% \caption{\small The average execution time of five tools.}
% \label{tab:efficiency}
% \end{table}

% \begin{table}[htbp]
% \centering
% \small
% \begin{tabular}{lccc}
% \toprule
% \textbf{Tool} & \textbf{Before} & \textbf{After} & \textbf{Improvement} \\
% \midrule
% analyze\_resource\_usage & 3656 & \textbf{526}  & 85.6\% $\downarrow$ \\
% analyze\_resort\_revenue & 3397 & \textbf{429} & 87.4\% $\downarrow$ \\
% analyze\_team\_scales & 1973 & \textbf{372}  & 81.1\% $\downarrow$ \\
% analyze\_associations & 1198 & \textbf{92} & 92.3\% $\downarrow$ \\
% analyze\_regional\_data & 827 & \textbf{97} & 88.3\% $\downarrow$ \\
% \bottomrule
% \end{tabular}
% \caption{\textbf{TODO}}
% \label{tab:efficiency}
% \end{table}

\subsection{Deployment Experience}
\label{sec:deployment}
An earlier version of VIVID has been deployed for three months in a city-level, state-owned tourism management company with more than 1,500 operational personnel. During this period, the system processed 2,375 orders. 
The order-management team changed from 15--20 personnel to three, while average per-order processing time decreased from ten minutes to under two minutes, including human-in-the-loop overhead. The observed final-task error rate was below 1\%\footnote{The final error rate is the proportion of processed orders with an erroneous outcome after human-in-the-loop operation.}. A voluntary and informed employee survey yielded 1,300 valid responses. In these aggregate self-reports, 96.92\% (1,260) reported improved work efficiency and 76.92\% (1,000) reported convenient access to multiple features. These results provide real-world evidence of VIVID’s operational use at scale, complementing the controlled evaluation. Appendix~\ref{app:deploy} describes the scenario setting in greater detail.

\section{Conclusion}
VIVID provides evidence for the operational viability of a complete tool-augmented agent system under enterprise constraints. Current agents support useful automation within a governed execution, but stage-level errors limit strict end-to-end autonomy. The practical value of agents therefore depends on bounded autonomy rather than model capability alone. To mitigate errors across multiple agentic steps, the interactive interface allows user observation and intervention. The Application Control Gateway further preserves organizational control by authorizing executable actions independently of model generation. To maintain responsiveness across the complete execution path, VIVID combines stage-wise KV-cache eviction with runtime monitoring and targeted optimization of slow business services. The deployment experience indicates that organizational control, human oversight, and full-path agent engineering must operate jointly under current limits in agent capability.

\section*{Limitations}
The deployed tool inventory reflects existing enterprise services and consequently includes some low-level operations that are shared across multiple task-specific tools, such as database access. This service-oriented organization may introduce redundancy in tool schemas and complicate tool discovery and maintenance. A hierarchical abstraction of shared operations could provide a more structured organization, but its effectiveness remains to be evaluated. The evaluation covers the deployed inventory within a deliberately narrow scope that primarily includes requests terminating in one business-tool invocation. Ambiguous requests and longer multi-tool trajectories require further exploration and stronger control over execution chains. Future evaluation could further examine human-in-the-loop interaction, including intervention patterns and their effects on task outcomes, as well as authorization robustness under adversarial conditions.

%The observed changes in staffing and processing time are based on deployment observations and may also reflect concurrent organizational changes; longer-term evaluations would provide stronger evidence of the system impact.

\section*{Ethical Considerations}
The employee survey was voluntary and conducted with informed participation, and its results are reported only in aggregate form. Training and evaluation data were anonymized and kept strictly separate from the production environment. For operational safety, authorization is enforced independently of model-generated arguments, while human control is retained over consequential actions.

%The observed staffing change raises potential workforce concerns. We report it as organizational context and do not treat workforce reduction as a design objective or a causal effect established by this study. Deployments of similar systems should involve affected employees, define accountability for automated actions, and monitor how automation changes job responsibilities and access to organizational resources.

\bibliography{custom}

\appendix
\clearpage
\renewcommand{\thefigure}{A.\arabic{figure}}
\setcounter{figure}{0}

\renewcommand{\thetable}{A.\arabic{table}}
\setcounter{table}{0}

\section{Interactive User Interface Interpretation}
\label{app:front}
\begin{figure}[ht]
    \centering
    \includegraphics[width=0.48\textwidth]{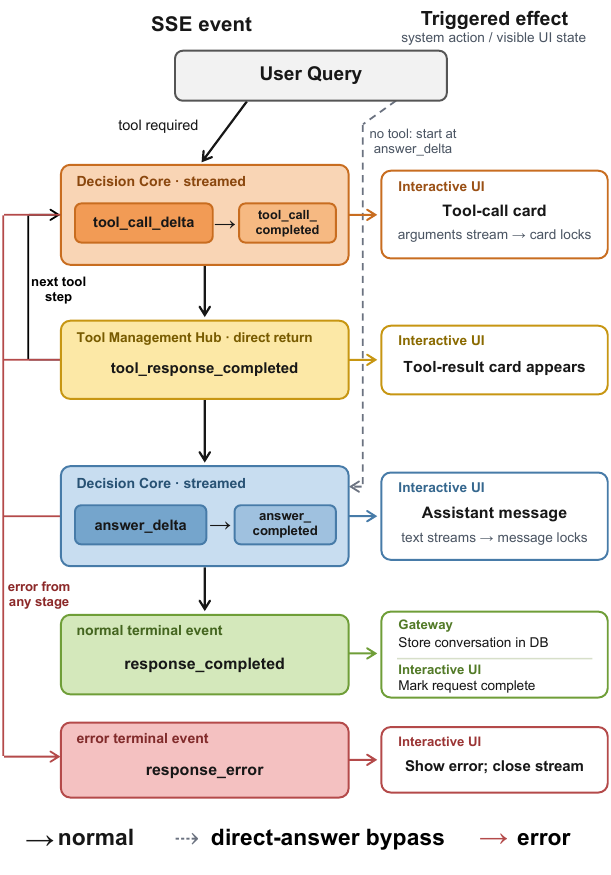}
    \caption{\small SSE event lifecycle and the corresponding system actions and user-visible interface states. Gateway means Application Control Gateway; Interactive UI stands for Interactive User Interface.}
    \label{fig:app-sse}
\end{figure}
\begin{figure*}[t]
    \centering
    \includegraphics[width=\textwidth]{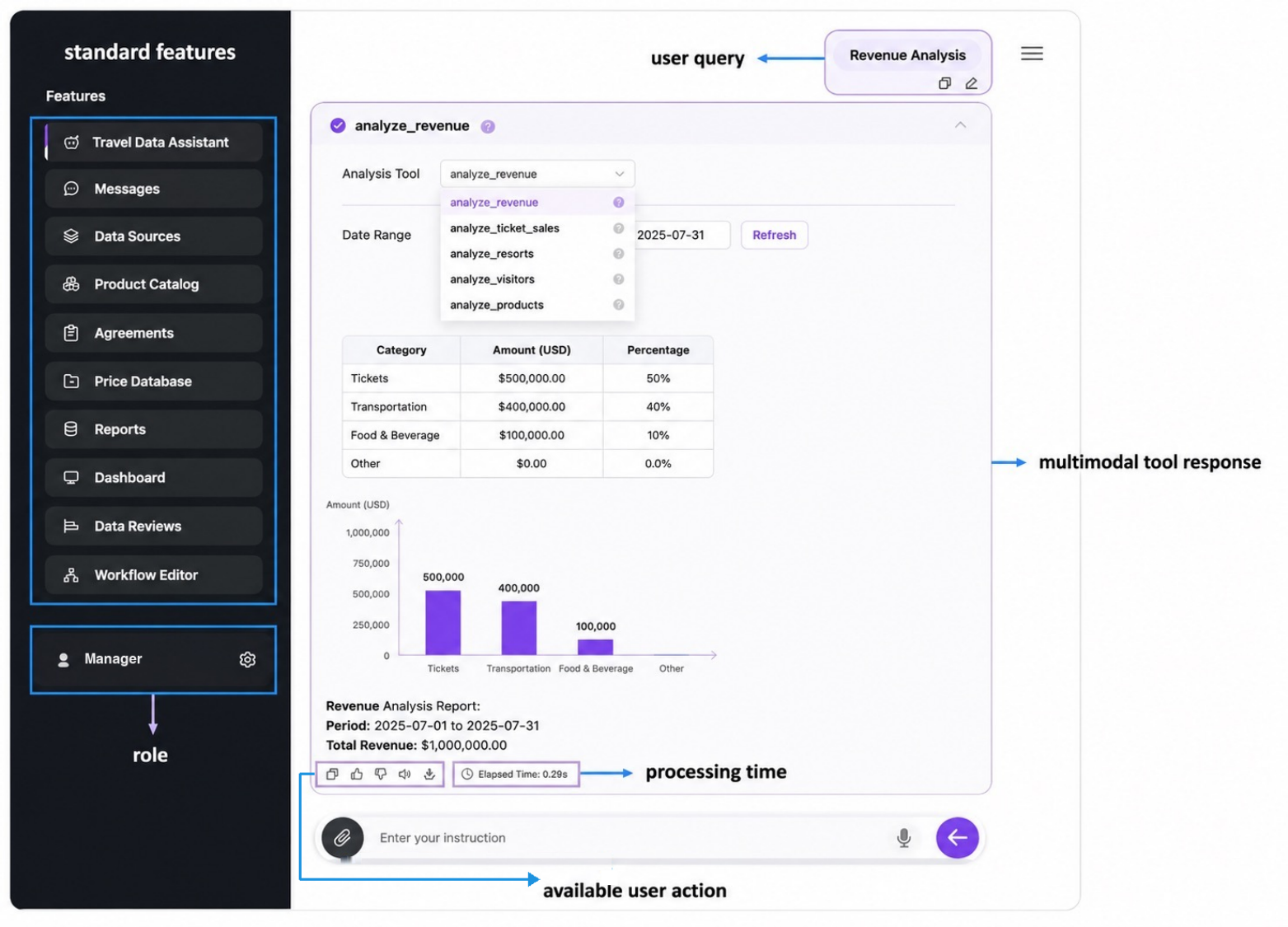}
    \caption{\small The prototype of the main agent.}
    \label{fig:app-prototype}
\end{figure*}

\paragraph{User Intervention}
Agent systems involve multi-step interactions with external tools, which introduce uncertainty compared with conventional chat systems. To improve transparency and user controllability, VIVID exposes intermediate execution states and supports human intervention throughout the process. Figure~\ref{fig:app-prototype} presents the main agent interface, which displays user queries, multimodal tool outputs, and elapsed execution time to provide visibility into the ongoing process. Users can override the automatically selected tool by choosing an alternative from the top-five candidates ranked by semantic similarity and can adjust task-specific parameters such as date ranges. System-managed information, including authorization context, is injected by the Application Control Gateway after access validation and remains inaccessible to both users and LLM generation. The interface also supports response regeneration and allows an active execution to be terminated when intermediate actions or results do not meet user expectations.

\paragraph{SSE Communication}
To support transparent interaction, VIVID maps the agent execution lifecycle into a sequence of Server-Sent Events (SSE), enabling incremental synchronization between the Decision Core and the user interface. As illustrated in Figure~\ref{fig:app-sse}, a tool-use request is represented through four execution stages. During tool preparation, \texttt{tool\_call\_delta} events progressively populate the tool-call card, while \texttt{tool\_call\_completed} finalizes the invocation. The Tool Management Hub subsequently executes the selected tool and returns \texttt{tool\_response\_completed}, which updates the corresponding result card\footnote{The integrated business services currently use request--response interfaces, so VIVID emits tool outputs after execution completes. Tools that expose streaming outputs could be supported through incremental tool-response events followed by
\texttt{tool\_response\_completed}.}. If additional actions are required, subsequent tool-call and tool-response events support multi-step execution. After tool execution completes, the Decision Core streams generated responses through \texttt{answer\_delta}, with \texttt{answer\_completed} marking response completion. Finally, \texttt{response\_completed} persists the interaction history and closes the current interaction. Requests without tool invocation directly enter the response generation stage, while failures during execution produce a terminal \texttt{response\_error} event for failure states.

\paragraph{Structured Response Rendering}
While SSE determines when interface updates occur, structured response rendering determines how heterogeneous execution results are presented to users. VIVID separates machine-oriented outputs from user-oriented representations: tool responses preserve structured data for downstream processing and persistence, while optionally providing display-oriented Markdown components mapped to interface elements. For example, analytical tools can render summaries or tables, navigation tools can generate actionable buttons, and visual tools can directly display generated content. This design enables the same execution results to support both continued agent processing and intuitive user interaction, reducing the cognitive load required for users to interpret and act on agent outputs.

\section{Initial Tool Development Exploration}
\label{app:tool}
\paragraph{Manual Creation Exploration}
In the early stage, we manually developed tools based on existing user queries. We observed that user queries were generally well specified, while the primary challenge arose from inconsistent tool schemas created by different engineers, each following their own conventions. To improve schema consistency, we established a unified specification for manual tool development. The main rules are summarized as follows:

\begin{itemize}
    \item Tool descriptions should clearly specify the provided business capability and expected output, while input parameters should define their semantics, types, valid ranges, and enumerated values when applicable;
    \item Tools associated with the same business entity should follow consistent categorization, and responses should adopt stable structures with explicit data-availability indicators.
\end{itemize}

These guidelines improved schema consistency and interpretability, but manual compliance verification remained labor-intensive. Moreover, a gap remained between developer-oriented tool definitions and user-facing interactions: exposing implementation-oriented function names (e.g., names following software engineering conventions with underscores) can introduce unnecessary complexity for end users. Therefore, we introduced user-oriented natural language names to the interactive user interface while preserving developer-oriented identifiers for LLM-based tool selection. These observations motivated our initial investigation into automated tool development.

\paragraph{Automatic Creation Exploration}
We designed a schema-guided tool development pipeline for generating application-specific tools with three stages using Claude Sonnet 4.5. In the first \emph{schema-design} stage, the LLM generator constructs tool specifications from either existing business API architectures, by identifying frequently used functions suitable for agent invocation, or user queries that reveal potential new tool requirements. This stage bridges existing service implementations and user-oriented tasks by producing structured schemas containing tool descriptions and argument definitions. In our preliminary evaluation, each generation batch produced 20 candidate tools, among which approximately 75\% were manually verified to align with real business scenarios across multiple batches.

The subsequent stages follow a standard software engineering process with a bottom-up generation strategy. In the second \emph{code-generation} stage, the generated schema is progressively transformed into three implementation layers: the Schema layer defines structured request and response models through Pydantic; the Service layer implements the underlying business logic; and the Controller layer exposes the functionality through authenticated FastAPI endpoints with standardized responses and error handling. This ordering follows the dependency hierarchy of software components, where service logic and external interfaces rely on well-defined data structures. Finally, the generated tools undergo \emph{unit testing} to verify code correctness and interface consistency. Failed implementations are returned to the code-generation stage for targeted correction, forming a constrained repair loop focused on implementation validity.

Although this exploration demonstrates the feasibility of schema-guided tool generation, fully LLM-driven development remains unsuitable as the primary production strategy. Enterprise deployment requires strict control over generated code, business logic, and data exposure, which is difficult to guarantee through autonomous generation alone. Enterprise tool ecosystems also evolve incrementally as new business requirements emerge. VIVID therefore adopts a semi-automatic production approach in which frequently used enterprise services are transformed into standardized tools through controlled development workflows. The automated pipeline remains a promising direction for accelerating initial tool creation and assisting developers in expanding tool collections.

\section{Preliminary Experiments}
\label{app:initial}
\paragraph{Embedding Model Exploration}
Initial supervised fine-tuning (SFT) increased Hit@5 by a 17.69\% relative improvement. Retrieval nevertheless remained the primary source of end-to-end failure, suggesting that the initial embedding model did not adequately align user expressions with tool intent. We therefore examined both task-specific embedding adaptation and the use of GPT-5 to normalize user queries and tool descriptions before retrieval. Qwen3-Embedding-0.6B (Qwen3-Emb)~\cite{qwen3embedding} was selected to balance retrieval quality with memory and latency requirements. The study was restricted to examples whose execution paths involved tool retrieval. During training, negative sampling generated one positive and five negative samples.

Table~\ref{tab:app-embedding} shows that the adapted embedding model achieved its highest Hit@5, 92.07\%, when both queries and tool descriptions retained their original wording. Rewriting either side reduced retrieval effectiveness, while rewriting both produced the lowest score of 82.07\%. Manual inspection attributed this degradation to semantic drift in the generated text. For example, GPT-5 transformed \textit{analyze the order status distribution} into the unnatural expression \textit{the \textbf{health degree} of the status} and occasionally introduced details absent from short queries. We consequently fine-tuned Qwen3-Embedding-0.6B on the original query--tool pairs. It achieved a 35.4\% relative improvement over bge-small-zh~\cite{bgeembedding} and was subsequently integrated into the main-agent workflow.

\begin{table}[t]
    \centering
    \footnotesize
    \setlength{\tabcolsep}{2.5pt}
    \begin{tabular}{@{}llcc@{}}
        \toprule
        & & \multicolumn{2}{c}{\textbf{Tool Description}} \\
        \cmidrule(l){3-4}
        \textbf{Embedding Model} & \textbf{Query}
            & \textbf{Original} & \textbf{Rewritten} \\
        \midrule
        Qwen3-Emb & Original
            & \textbf{92.07} & 86.90 \\
                        & Rewritten
            & 87.59 & 82.07 \\
        \midrule
        bge-small-zh    & Original
            & 68.00 & -- \\
        \bottomrule
    \end{tabular}
    \caption{\small Hit@5 (\%) for different query and tool-description representations.}
    \label{tab:app-embedding}
\end{table}

\paragraph{Pre-audit Diagnostic Comparison}
Table~\ref{tab:app-history} summarizes the diagnostic configurations evaluated before the final dataset audit. Successful Hit@5 computation for the untuned backbone means its success on the first step. The remaining metrics require a valid follow-up business-tool call, which the backbone does not produce reliably. Unlike one-step function calling, this transition requires observation-conditioned continuation under domain understanding, for which the untuned model has not received task-specific trajectory supervision. It instead frequently generates repetitive text, incomplete responses, or malformed actions after receiving the retrieval result. Selection accuracy, argument accuracy, and end-to-end success are therefore omitted. This behavior is consistent with evidence that single-turn function-calling ability does not ensure reliable stateful, multi-step execution~\cite{patil2025berkeley}, motivating the subsequent trajectory-level supervised adaptation~\cite{liu2025toolace}.

Adapting the retriever with SFT substantially improves both retrieval quality and end-to-end performance, raising Hit@5 from 72.97\% to 89.75\% and success rate from 45.58\% to 55.83\%. The corresponding decrease in selection accuracy does not indicate degraded selection performance, as improved retrieval exposes the selector to previously missed and more challenging queries, lowering its conditional accuracy. Thus, selection accuracy should be interpreted jointly with retrieval coverage and end-to-end success.

\begin{table}[ht]
    \centering
    \footnotesize
    \setlength{\tabcolsep}{1pt}
    \begin{tabular}{lcccc}
        \toprule
        \textbf{Configuration} & \textbf{Hit@5} &
        \textbf{Select. Acc.} & \textbf{Arg. Acc.} &
        \textbf{E2E SR} \\
        \midrule
        Backbone
            & 62.00\% & -- & -- & -- \\
        SFT
            & 72.97\% & \textbf{87.04\%} & 71.76\% & 45.58\% \\
        + retriever
            & \textbf{89.75\%} & 78.35\% & \textbf{79.40\%} & \textbf{55.83\%} \\
        \bottomrule
    \end{tabular}
    \caption{\small Pre-audit diagnostic results on the original evaluation snapshot. Selection and argument accuracy are conditional on successful completion of their preceding stages. \textbf{--} indicates that the untuned model did not reliably proceed to the corresponding stage.}
    \label{tab:app-history}
\end{table}

Manual error analysis subsequently identifies an evaluation artifact: several constructed queries admitted multiple functionally valid tools. A single-reference protocol classifies such alternatives as model errors even when they satisfy the expressed intent, a recognized difficulty in evaluating agents with multiple valid execution paths~\cite{gu2024blade}. Following the human-resolvability criterion motivated by conversational grounding~\cite{clark1991grounding}, we restrict the feasibility evaluation to queries for which manual review can establish a uniquely defensible reference tool. Task adaptation uses 1,601 training examples.

The final audited snapshot contains 283 queries that are disjoint from these training examples, with both query membership and query--tool mappings revised from the pre-audit snapshot to remove ambiguous queries and incorporate more challenging ones. Final results in the main paper are reported on this final snapshot. This scope isolates the initial viability of reliable single-tool execution, while ambiguous requests and multi-tool trajectories require clarification-aware or multi-reference evaluation~\cite{zhang2025clarify} and are therefore left for future study.

\section{Development-Time Monitoring}
\label{app:monitoring-diagnosis}
To complement runtime monitoring in production evaluation, we present a development-time case study demonstrating how the monitoring system supports performance diagnosis and optimization. The collected measurements target specific interfaces and workloads during iterative debugging and illustrate the diagnostic workflow. The monitoring system captures interface-level latency distributions, request failures, call-chain traces, and function-level profiles. An initial analysis of 186 \emph{Chat} requests revealed a median latency of 4.80\,s, with P95 and P99 latencies of 13.80\,s and 26.78\,s, respectively. Call-chain analysis showed that over 83\% of requests shared the same dominant invocation pattern, and over 85\% involved ten recorded operations. These results indicated that repeated interactions and insufficient concurrency caused the long-tail latency.

Based on these observations, we performed controlled load tests within \emph{Chat} to evaluate two targeted optimizations. First, a JMeter workload with 50 client threads, a 5-second ramp-up period, and 10 iterations per thread generated 500 \emph{List} requests. Under the original configuration, 2.6\% of requests failed because persistent connections expired before reuse. Increasing the Uvicorn keep-alive timeout from 5\,s to 15\,s eliminated these failures under the same workload without affecting average latency, improving connection stability.

Second, function-level profiling identified tool-list construction as a major latency contributor: an initial \emph{List} request required 1.290 seconds, of which 1.167 seconds were spent constructing the response. In particular, initializing the third-party search manager accounted for 1.134 seconds (97.1\%) of this overhead. Although subsequent requests benefited from singleton and LRU caches, the endpoint still reconstructed the complete response object for each invocation. Since registered tools evolve infrequently, we introduced endpoint-response caching to store the fully constructed \emph{List} response in memory and directly serve subsequent requests. Under a workload of 50 threads, a 5-second ramp-up period, and 100 iterations per thread (5,000 requests in total), this optimization reduced mean latency from 1,288\,ms to 24\,ms and P95 latency from 3,774\,ms to 44\,ms. Throughput increased from 29.1 to 686.8 requests per second, while the failure rate decreased from 0.12\% to zero.

\section{Deployment Details}
\label{app:deploy}
The tourism management enterprise manages seven major scenic areas and twelve business categories spanning land, water, and air services. Its operational work includes highly non-standardized requests that cross conventional application boundaries. VIVID provides a unified conversational entry point to the existing enterprise software and records execution data and errors for subsequent system maintenance.

\end{document}